\documentclass[preprint2,final]{aastex}
\usepackage{natbib}
\usepackage{float}
\usepackage{sidecap}
\shorttitle{Bulge Variables}
\shortauthors{Nataf et al.}
\begin{document}

\title{Finding the Brightest Galactic Bulge Microlensing Events with a Small Aperture Telescope and Image Subtraction}
\author
    {
      D. M. Nataf\altaffilmark{1},
      K.~Z. Stanek\altaffilmark{1},
      G.\'A.~Bakos\altaffilmark{2},
    }
    \altaffiltext{1}{Department of Astronomy, Ohio State University, 140 W. 18th Ave., Columbus, OH 43210}
    \altaffiltext{2}{NSF Fellow, Harvard-Smithsonian Center for Astrophysics, 60 Garden St., Cambridge, MA~02138}
%\author{D. M.~Nataf, K.~Z.~Stanek, G.\'A.~Bakos\altaffilmark{1}}
%\affil{Department of Astronomy, Ohio State University, 140 W. 18th Ave., Columbus, OH 43210,  Harvard-Smithsonian Center for Astrophysics, 60 Garden St., Cambridge, MA~02138}
\email{nataf@astronomy.ohio-state.edu, kstanek@astronomy.ohio-state.edu, gbakos@cfa.harvard.edu}

\begin{abstract}Following the suggestion of \citet{1998GouldDepoy} we investigate the
feasibility of studying the brightest microlensing events towards the
Galactic bulge using a small aperture ($\sim$10$\;$cm) telescope. We used
one of the HAT telescopes to obtain 151 exposures spanning 88 nights
in 2005 of an $8.4^\circ \times8.4^\circ$ FOV centered on (l,b) =
(2.85, $-$5.00). We reduced the data using image subtraction
software. We find that such a search method can effectively contribute
to monitoring bright microlensing events, as was advocated. Comparing
this search method to the existing ones we find a dedicated bulge
photometric survey of this nature would fulfill a significant niche at
excellent performance and rather low cost. We obtain matches to 7
microlensing events listed in the 2005 OGLE archives. We find several
other light curves whose fits closely resemble microlensing
events. Unsurprisingly, many periodic variables and miscellaneous
variables are also detected in our data, and we estimate approximately
50\% of these are new discoveries. We conclude by briefly proposing Small Aperture Microlensing Survey, which would monitor the
Galactic bulge around the clock to provide dense coverage of the
highest magnification microlensing events.
\end{abstract} 
\keywords{binaries: eclipsing - Galaxy: bulge - gravitational lensing - techniques: photometric}
%Cepheids – galaxies: distances and redshifts – galaxies: individual (M100, NGC 4321) – stars: early type 

\section{Introduction}

Constraints on MACHOs \citep{1993Bennett}. Observation of a binary lens \citep{1994Udalski}. Spectroscopic profiling of
faint stars \citep{1996Lennon}.   Direct measurements of stellar masses
\citep{2002An}. Extrasolar planets \citep{2004Bond}. Terrestrial
parallax measurements of a thick-disk brown dwarf \citep{2009Gould}.
These are just some of the ``firsts'' in the long string of successes
that have followed since \citet{1986Paczynski} got the field of
microlensing started with a ``futuristic'' plan to probe the Milky
Way's halo.

But even with all these successes, there is still ample discovery
space remaining in the field of Galactic microlensing. Exploring an
unfulfilled niche, \citet{1998GouldDepoy} advocated using a small
aperture telescope ($\sim$65$\;$mm diameter), a $6^\circ \times6^\circ$
field of view, and difference imaging to study microlensing events
towards the Galactic bulge. Such a survey would allow one to observe
objects at higher brightness and to observe the entire bulge with a
few pointings. Three key scientific advantages were specified: the
survey would provide an analog to microlensing studies of M31; it
would yield a complete inventory of bright bulge variables; and data
could be taken at the peaks of extreme magnification microlensing
events.

\begin{table*}[t]

\caption{Observational summary of major current and future
microlensing surveys. Limiting magnitudes are for 10 \% photometric
precision, except for MPF where the number cited is for 1 \%
precision. Surveys that are line-separated are proposed future surveys
(KMTNet is already funded).}

\scalebox{0.95}
{\begin{tabular}{ l c l l l l l l l}
\hline\hline
Survey & Aperture & Pixels & FOV & Limiting & Saturation & Cadence & Bulge \\
&(m) & & Square Degrees & Magnitude & Limit & (Approximate) & Fields \\
\hline
OGLE-III & 1.3 & 8192x8192 & 0.58x0.58 & I~19.5 & I~12 & 1/(3 nights) & 267 \\
MOA-Cam3 & 1.8 & 10x(2Kx4K) & 1.32x1.65 & I~18.5  & I~12.5 & 5/night & 22  \\ %[1ex]
HAT 2005 Run & 0.11 & 2048x2048 & 8.4x8.4 & I~12.5 & I~8 & 1.5/night & 1 \\
\hline 
KMTnet & 3 x 1.6 & 4x(10Kx10K) & 2x2 & I 21 & I 12 & 144/(24 hours) &  4\\
\hline\
MPF & 1.1 & $\sim$8000x6000 & 0.95x0.65 & J 20.5 & -- & 150/(24 hours) & 4  \\ 
\hline
SAMS & 3$\times$0.1 & 4096x4096 & 11x11 & I 14.5 & I 8.0 & 300/(24 hours) & 1 \\
\hline
\end{tabular}
\label{table:surveys}
}
\end{table*}

Progress within the field of microlensing along with technological
advances have modified the prospects for some of these advantages but
not eliminated them. As seen in Table \ref{table:surveys}, there
remains a place for a survey observing the entire Galactic bulge in
one or few pointings, at high brightness and high cadence. We obtain
the specifications for OGLE, MOA and MPF from \citet{2003Udalski},
\citet{2008Sako}, \citet{2000Bennett} and \citet{2007Bennett}. The
preliminary specifications for KMTNet, which will network observations
at three different locations for round-the-clock time-series, came
from \citet{2009Gaudi2}, Park
(2008)\footnote{http://www.astronomy.ohio-state.edu/$\sim$microfun/MF1/Talks/Park$\_$KMTNet.pdf},
and Han (private communication).

Surveys of microlensing in M31 are building on the pioneering work of \citet{1996Crotts}. The Angstrom project has been observing since 2005, and has had its early warning system operational since 2007 \citep{2007Darnley}. \citet{2009Novati} have also conducted 2 years of observations of M31. Closer to home, galactic bulge variability surveys have often either been limited in FOV such as the Spitzer-IRAC survey \citep{2008Ramirez} or been optimized to a different range of magnitudes such as OGLE \citep{2008Soszynski}. 

Meanwhile, in the case of high-magnification microlensing events,
networks of amateur observers such as MicroFUN \citep{2008Gould} refine time series at
the peaks, but it is hard to maintain the reliability and consistency
that would be obtained from a dedicated and automated
survey. Nevertheless, high-magnification microlensing events have
blossomed into a very critical subfield of microlensing. In his recent review
of the field, \citet{2009Goulda} mentions 7 planets already discovered
by this method as well as 6 more whose details are not yet
published. The 4th microlensing planet, OGLE-2005-BLG-169, was
constrained by observer Doekunn An taking over 1000 observations at
the magnification peak after being told of the urgency \textit{the
night of} the event \citep{2006Gaudi}. The 5th and 6th microlensing
planets, OGLE-2006-BLG-109, were perhaps the first Jupiter-Saturn
analog found, and the first multiplanet system found using
microlensing. The detail extracted from this event was possible due to
observations being taken by 12 observatories at or near the peak
\citep{2008Gaudi}. It is unfortunate that such good observational
support is not currently available to all high-magnification
microlensing events. The case for a dedicated, automated and
high-cadence survey is evident. 

\citet{1998GouldDepoy} also advocated a broad ($\sim30\arcsec$) PSF. Such a PSF is much less dependent on atmospheric conditions, and would reduce pixelization noise, which falls as the inverse 4th
power of the PSF \citep{1996Gould}. This end can now be achieved by
software methods rather than hardware methods. For example, the HAT
telescopes use a sequence of small pointing steps during the exposure
to broaden the PSF \citep{2004Bakos}. Another technique employed to
broaden the PSF, is to use charge shifting by means of an orthogonal transfer
array \citep{2009Johnson}.

Motivated by the proposal of \citet{1998GouldDepoy} and by our
results, we encourage development of a Small Aperture Microlensing
Survey (SAMS). A generic list of technical and observation
specifications of such a survey is included in Table
\ref{table:surveys}.

In this paper we report on the performance of our feasibility
study. In Section \ref{section:Data} we provide a synopsis of our
observational data set. We summarize our reduction methods in Section
\ref{section:DR}.  In Section \ref{section:MA} we report on our search
for both known microlensing events as well as previously undiscovered
ones. In Section \ref{section:OVS} we comment on the potential to find
new variable stars within the bulge, presenting a few examples. We
conclude in Section \ref{section:Conclusion}.

\clearpage
\begin{figure*}[p]
\begin{center}
\includegraphics[trim = 0mm 14mm 0mm 0mm, clip, totalheight=0.81\textheight]{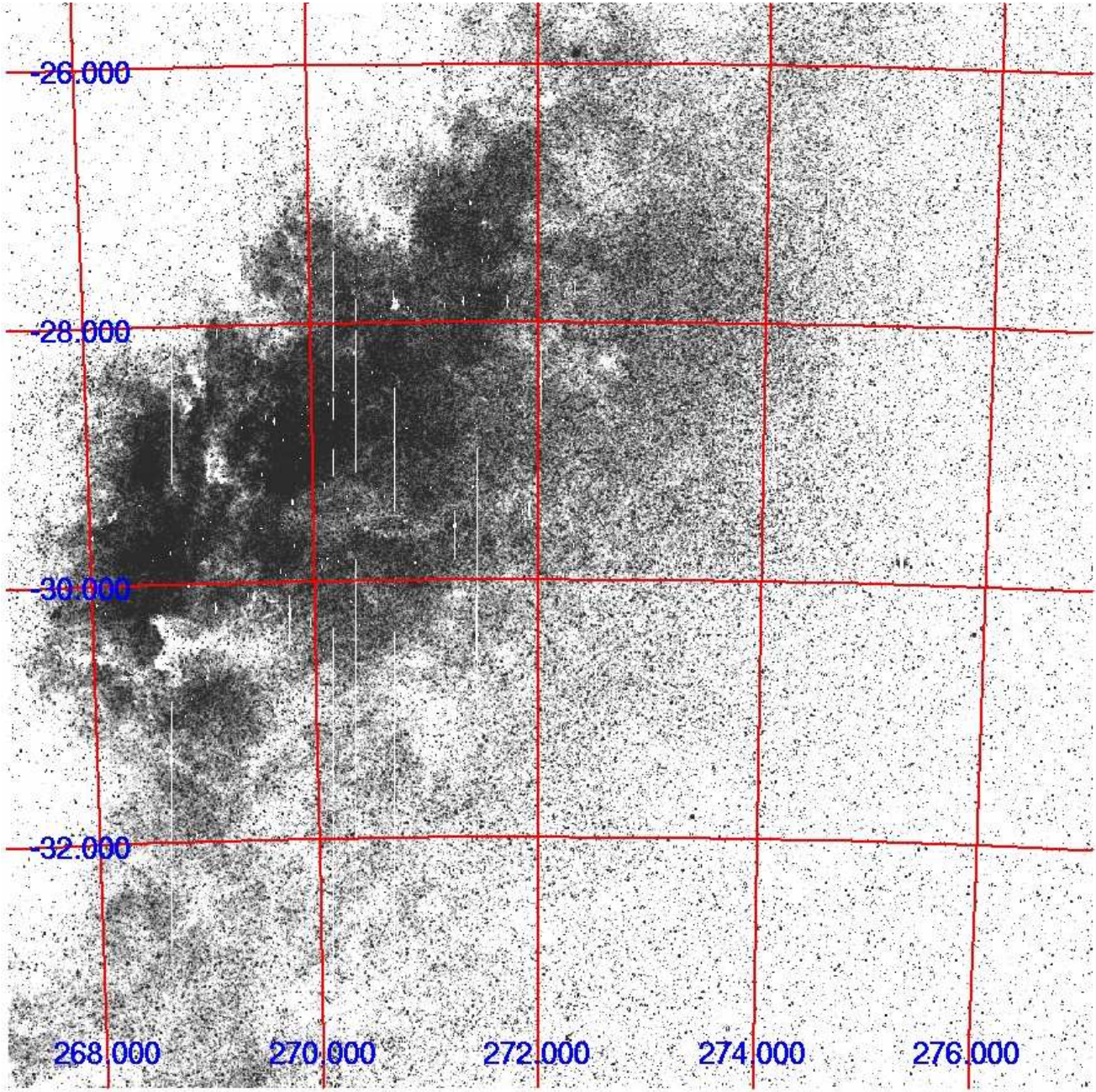}
\end{center}
\caption{Image of our field, with $8.4^{\circ}\times 8.4^{\circ}$~FOV
centered on (l,b) = (2.85, $-$5.00).  RA and DEC are shown in blue.}
\label{ReferenceImage}
\end{figure*} 
\clearpage

\section{Data}
\label{section:Data}

We used one of the HAT telescopes, HAT-9, located on Mauna Kea,
Hawaii. A Canon 11~cm diameter f/1.8L lens was used to image onto an
Apogee AP10 2K$\times$2K CCD. The resulting pixel scale is
$14\arcsec$. A detailed description of the equipment and observing
program can be found in \citet{2004Bakos}.

Our observations were taken over an 88 day span between July 22nd and
October 18th, 2005, yielding 151 exposures over 61 distinct
nights. Our $8.4^{\circ}\times 8.4^{\circ}$~FOV is centered on
$(\alpha,\delta)$=(18:11:54, -28:57:28) (J2000.0), in the general
direction of Baade's window, see Figure \ref{ReferenceImage}. Our
photometric precision ranges from 1\% for the brightest stars (I
$\sim$ 9) down to 10\% at the faint end (I $\sim$ 13). Thus, our
feasibility study achieved the wide field and the magnitude range
necessary to gauge the viability of a dedicated and automated survey,
but we did not implemented the high-cadence.

\section{Data Reduction}
\label{section:DR}

Preliminary CCD reductions such as dark current subtraction and
flat-fielding are based on IRAF and discussed in \citet{2004Bakos}. In
subsequent reductions we largely follow the procedure described in
\citet{2004Hartman}, which also dealt with HAT data for a dense
stellar field. Key parts and specific changes are discussed below.

\subsection{Image Subtraction and Photometry}

Due to our 2K$\times$2K CCD and $\sim10^5$ sources, we expect a
typical source to be within $\sim$7 pixels of another source. However, as
can be seen from our reference image (Figure \ref{ReferenceImage}),
sources are not evenly distributed and thus we are deep within the
domain of dense-field photometry. Additionally, there will be many
unresolved stellar sources contributing flux and some sources
contributing varying flux throughout our image. To subtract the
non-variable components of the images and to obtain light curves of
variable objects we use the ISIS package, see \citet{1998AlardLupton}
and \citet{2000Alard}, by now a standard tool when dealing with
crowded field variability studies. 

\begin{figure}[ht]
%\plotone{RMSvsMean_Magnitude}
\includegraphics[totalheight=0.39\textheight]{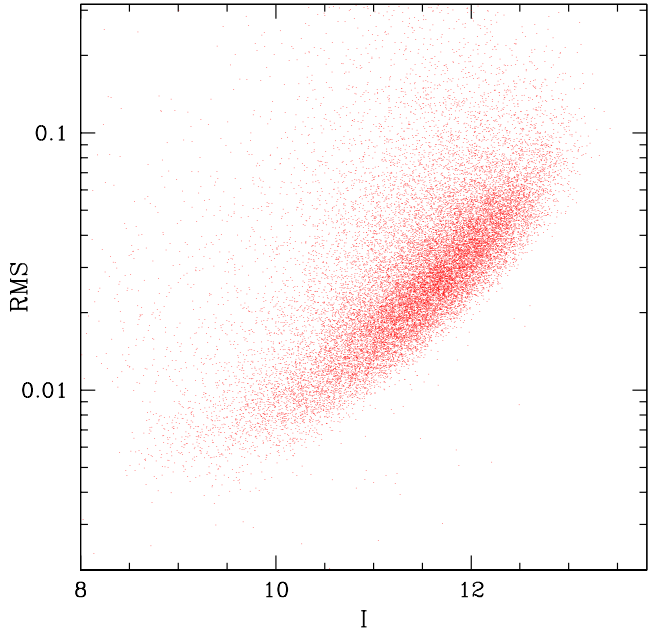}
\caption{RMS vs apparent $I$-band magnitude for a representative sample of points. There were few outliers left out of this plot.}
\label{RMSvsMean}
\end{figure}

Following the usual procedure
\citet{2004Hartman}, we used the image with the best spatial
resolution to perform the astrometric calibration step.  We then
constructed a reference image from 25 of our frames, selecting from
frames that had both sharper resolution as well as attempting to have
epochs spread throughout the time series. The first condition
facilitates reducing the impact of blending on our data, and the
second increases the probability that short-duration transient events
break our magnitude threshold and show up on succeeding source
lists. As the full width at half maximum (FWHM) was generally narrower
in the earlier part of the time-series, taking only the sharpest
images would have significantly limited the prospects for detecting
short-duration transients.

To construct the source list, we used the DAOphot/ALLSTAR package
\citep{1992Stetson}. Our photometric list is comprised of 115,624 objects  - many of them
likely to be blends of a number of stars. We then used ISIS to
produce the light curves. Figure \ref{RMSvsMean} shows the light curve
RMS vs. apparent magnitude for our sources.  A few key features can be
observed.  The \textquotedblleft main\textquotedblright sequence of
points, which is at approximately lowest RMS for a given magnitude,
corresponds to the non-variable sources. There is an intrinsic width
due to vignetting. A higher RMS is obtained from fainter sources, as
one would expect due to the inverse square-root scaling of photon
noise. There is also the impact of the surface brightness of the sky,
a significant issue with $14\arcsec$$\times$$14\arcsec$ per pixel and
when pointed towards the bulge. The sparser population on top with
much higher RMS are the candidate variable sources. At the bright end
of the distribution, a magnitude precision of ~4 millimagnitudes (with
5 minute cadence) is achieved, increasing to 0.1 magnitudes for the
fainter objects. This is a demonstration of the prospects for
time-series photometry in a dense, wide-field setting using image
subtraction. 

It is for these same brighter stars that a small-aperture
telescope would most contribute to the observational parameter
space. 7,714 of the sources have a precision below 10 millimagnitudes,
34,323 have a precision below 20 millimagnitudes and 91,218 have a
precision below 50 millimagnitudes.

Magnitude calibrations were done by comparing instrumental magnitudes
to photometry of OGLE objects in our field  with
I$\le$12.5, see \citep*{1997Udalski} and \citep{2005Szymanski}. We then applied a $15\arcsec$ matching radius to gauge the
magnitude shift. Using the median $\Delta$I for the magnitude shift
and the upper and lower quartiles to estimate our error, we obtained a
shift of (4.6 $\pm$ 0.2) magnitudes.

\section{Microlensing Analysis} 
\label{section:MA}

There are currently several groups actively searching for microlensing
events towards the Galactic bulge, notably OGLE, MOA, and the MicroFun
network. A cursory look at the list of partners in MicroFun \footnote{http://www.astronomy.ohio-state.edu/$\sim$microfun/} reveals
the use of equipment ranging in size from 0.25 meters (Perth
0.25-meter f/6.3 Telescope) to 2.4 meters (MDM observatory). At the
time of writing, science-grade microlensing detections had not been
cataloged from data obtained from a telescope whose primary is as
small as that of the HAT telescopes - 11cm.

We searched both for the signals of known microlensing events, and
also for events not previously discovered.

\subsection{Previously Known Microlensing Events}

We relied on the OGLE-III Early Warning system's catalog from 2005
\citep{2003Udalski}\footnote{OGLE-III Early Warning System URL:
http://ogle.astrouw.edu.pl/$\sim$ogle/ogle3/ews/ews.html}. 597 events
took place in 2005. We kept those that were within our field of view,
that had their peak magnitude within 30 days of our observation
window, and that had their peak magnitude above I $\sim$ 14.5. The 30
day time frame was a generous match for the typical crossing time, in
the hope we might catch the tail end of some events. A peak magnitude
of I $\sim$ 14.5 was used so that the minimum magnitude would yield a
signal, as magnitudes below that would wash into noise at our coarse
pixelization, low exposure times, and dense sky background. As OGLE's
magnitude-sensitivity range is typically between I $\sim$ 12-19, this
represented some of the brighter OGLE events.

\begin{figure*}[p]
\includegraphics[totalheight=0.8\textheight]{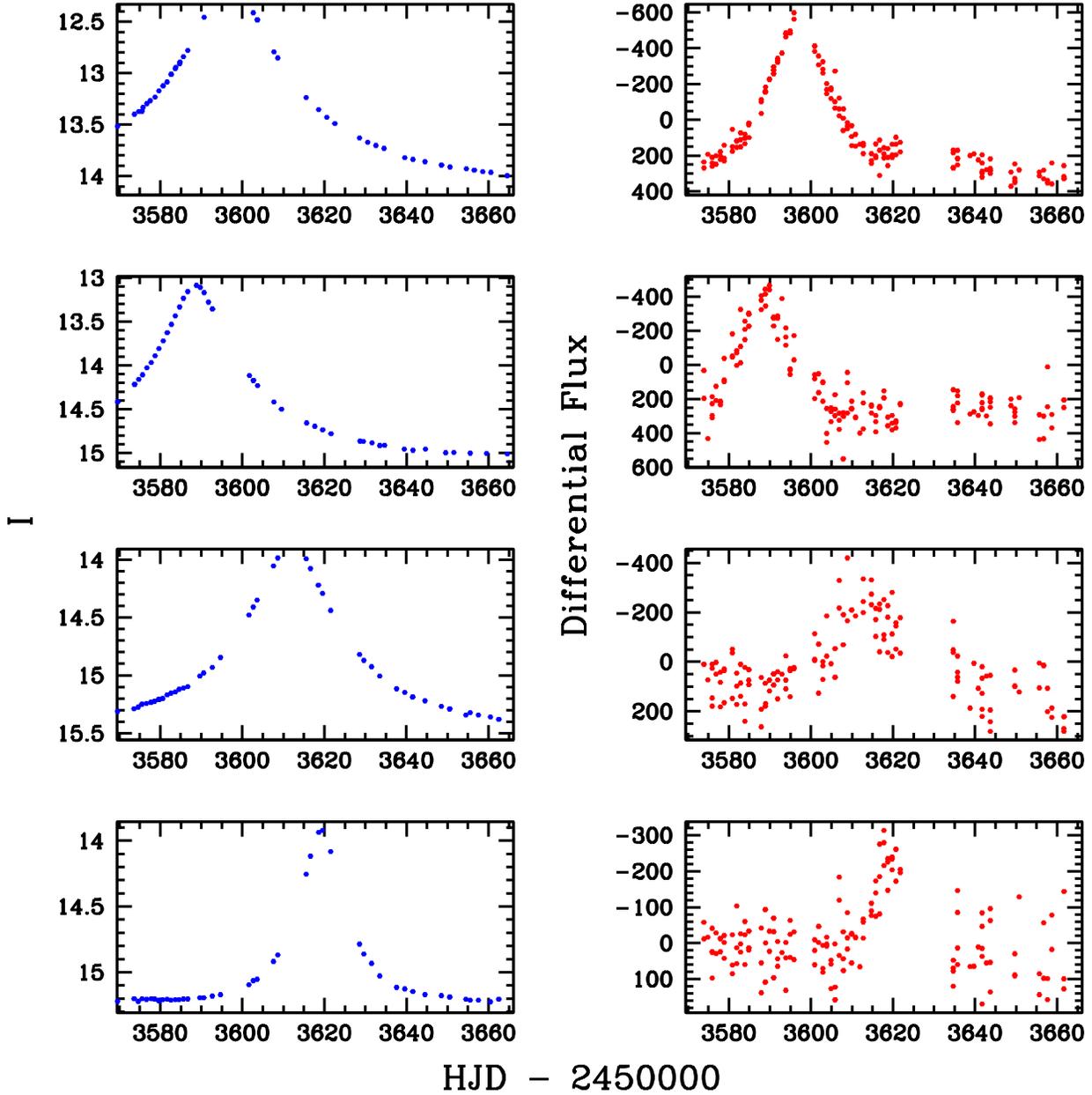}
\caption{Plot of known microlensing events with more visually
significant matches. The original OGLE light curves are plotted on the
left, and our matches are plotted on the right. Increase in flux displayed as a negative differential count by convention.}
\label{KME1}
\end{figure*}

\begin{figure*}[p]
\includegraphics[totalheight=0.8\textheight]{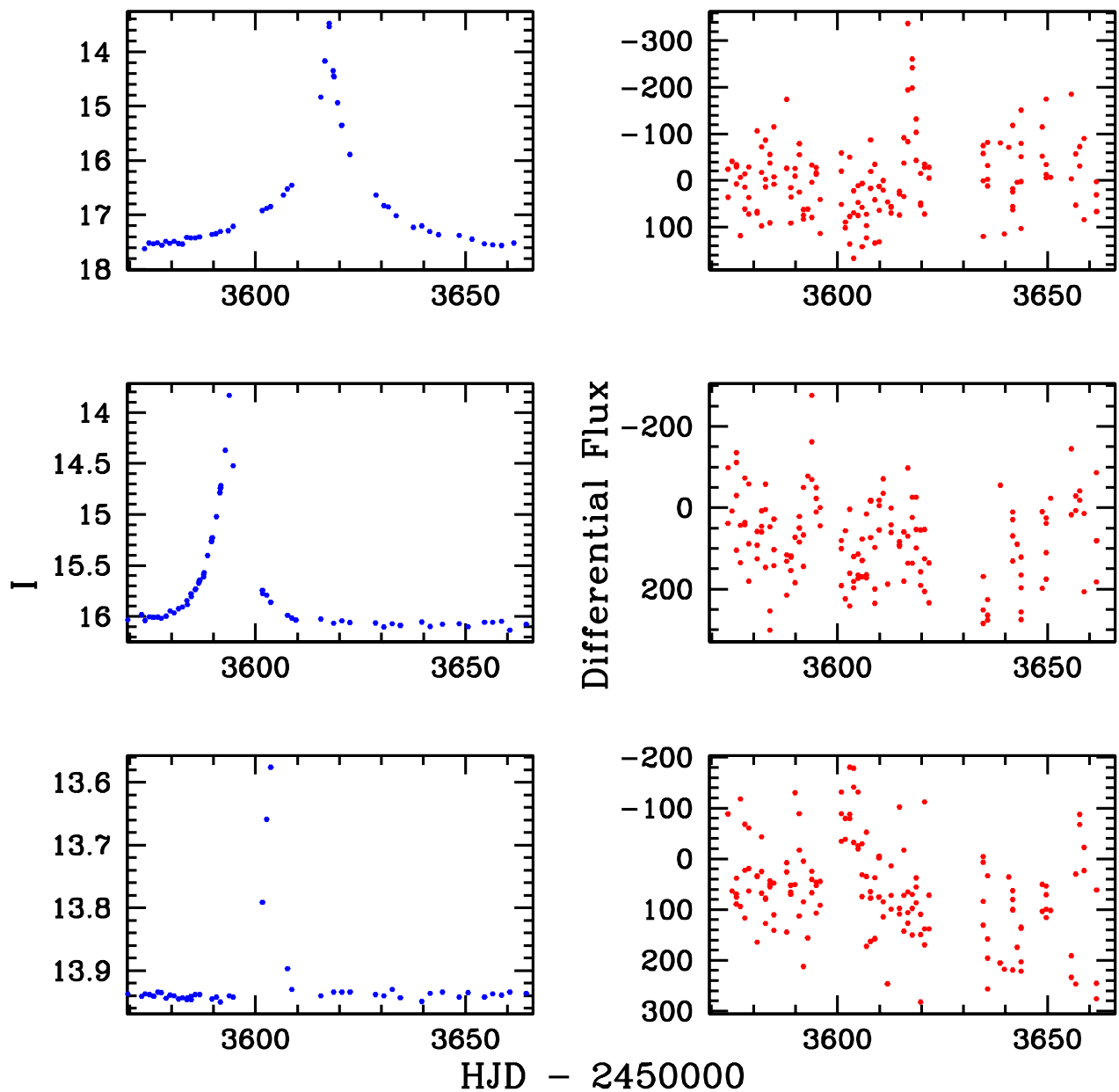}
\caption{As in Fig.3, this time with the three less good microlensing matches.}
\label{KME2}
\end{figure*}

That left us a list of 27 events. Plotting of the light curves for
sources centered at the OGLE coordinates yielded 7 (visual)
confirmations. Of these 3 are less clear (Figure \ref{KME2}), and 4
are of much higher quality (Figure \ref{KME1}). The visibly clearest
events are generally those with larger change in magnitude and higher
peak brightness. Due to our low cadence, events with a longer Einstein
crossing time were also easier to recognize.

Observational trends manifest themselves in the two figures. The
scatter in the plots drop substantially for the brighter events -
which is precisely where one would need lower scatter for an
accompanying survey. 

We failed to detect many events whose parameters
appear as those of the events we did detect or better. For example,
BLG-452 had its best fit OGLE curve peak at HJD 2453584.3, with a base
magnitude of ~19 and a magnification of ~8.7 magnitudes. However, the
brightest point actually recorded by OGLE corresponded to I = 18.002
due to the very small period of magnification, and the peak brightness
only reached by the fit for a period of hours. While we have three
observations on that night, they were within 30 minutes of one another
and $\sim$10 hours apart from the estimated time of peak brightness.

\clearpage
\clearpage

\clearpage
\clearpage

\subsection{Search for New Microlensing Events}

We investigated the possibility of additional microlensing events not
previously known. Two different methods were used. The first method
was to fit our existing light curves obtained from DAOphot to
microlensing parameters, and the second was to look for events in the
sources that did not show up in the DAOphot list but did show up in
the ISIS \texttt{abs.fits} image, a superposition frame of all the
absolute variations from the reference flux.

\subsubsection{Method 1: Using the DAOphot Source List}

Sources with a $J_{S}$ $\geq$ 2 were first selected, leaving a list of
9,775 light curves. This selection step immediately removed the
possibility of finding any short-duration ($t_e$ $\lesssim$ few days)
microlensing events.

With our narrow observational window and low cadence we would not have
been able to properly classify such events as microlensing events and
not, for example, as dwarf novae.

We fit single-lens microlensing parameters to the light curves using a
search and classification algorithm developed by Wyrzykowski
et. al. (2009, in preparation). We removed light curves with an
Einstein crossing time greater than 88 days those with a time of
maximum amplification was outside our observational window. We then
sorted the light curves by normalized $\chi^2$, and then manually
looked at the light curves. The best light curves were then refitted
with blending allowed to float as a free parameter.

\begin{figure}[ht]
%\plotone{CloseULensingCalls}
\includegraphics[totalheight=0.35\textheight]{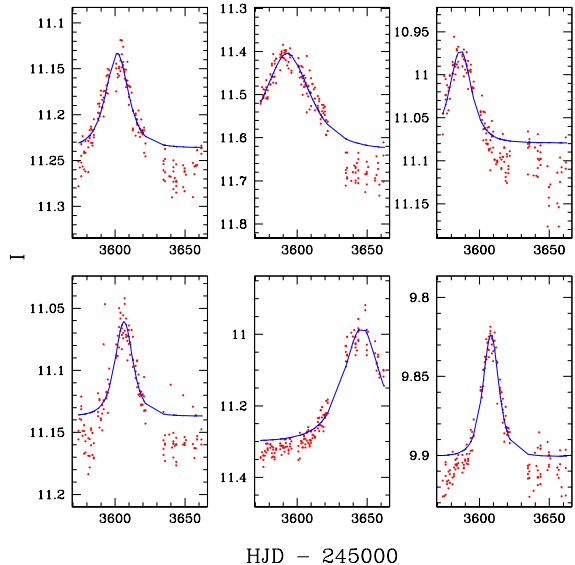}
\caption{Some of the light curves from our original source list which
came closest to being well-fit by microlensing.}
\label{CULC}
\end{figure}

A sample of the best light curves with their microlensing fits are
presented in Figure \ref{CULC}. There were no light curves that we
deemed to definitively be microlensing events. Our brief time window
does not allow us to comprehensively know which of the sources are
normally non-variable. Many of the inherent limitations would whither
away in a study with a longer baseline and higher cadence. Of the 7
OGLE microlensing events recognized within our data, only one (BLG
259) with $J_{s}$ of 3.19, would have made the cut of $J_{s}$
$\geqslant$ 2. It does not, however, show up in our source list.

The reduced $\chi^2$ for the 6 light curves selected in the plot
ranged from 1.72 to 2.24. There were some light curves with much lower
values, but these were not considered good fits as the numerical fits
came from spurious circumstance, \textit{i.e}. the time of maximum
amplification being outside our observation window.

\subsubsection{Method 2: Searching for Missed Sources}
The source list obtained from DAOphot will miss some light
sources. Sources that are faint with respect to the background and have
most or all of their brightening occur in frames which were not in the
reference image will not make the list. We compiled a list of sources
on the abs.fits image and removed those whose coordinate were within 1
pixel of sources already listed. 2,657 variable sources were added
this way.

These sources were fit to microlensing curves using a Perl script. As
only the differential flux was available, we modified the Paczy\'nski equations
for numerical convenience. Without blending and with knowledge of the base
flux one would use:

\begin{equation}
U(t) = \sqrt{u_{min}^2+(\frac{t-t_o}{t_e^2})^2}
\end{equation}
\begin{equation}
A(t) = \frac{U^2+2}{U\sqrt{U^2+4}}
\end{equation}

Removing the base flux and solving for the variable flux:
\begin{equation}
\Delta F=\Delta F_{max}G_{U_{min}}
\frac{U^2+2-U\sqrt{U^2+4}}{U\sqrt{U^2+4}}
\label{ModPac}
\end{equation}

Where:
\begin{equation}
G_{U_{min}}= \frac{u_{min}\sqrt{u_{min}^2+4}}{u_{min}^2+2-U_{min}\sqrt{u_{min}^2+4}}
\label{Ufactor}
\end{equation}

Equation \ref{ModPac} has the further advantage of being
blending-independent (assuming non-variable blending). We did find a
light curve corresponding to BLG-259, and with a $\chi^2$ = 1.11 it
had one of the best fits. The light curves with an even lower $\chi^2$
were either spurious \textit{i.e} with huge errors or fewer data
points, or they had their peak at the edge of our observational
window.

\begin{figure}[ht]
\begin{center}
\includegraphics[totalheight=0.37\textheight]{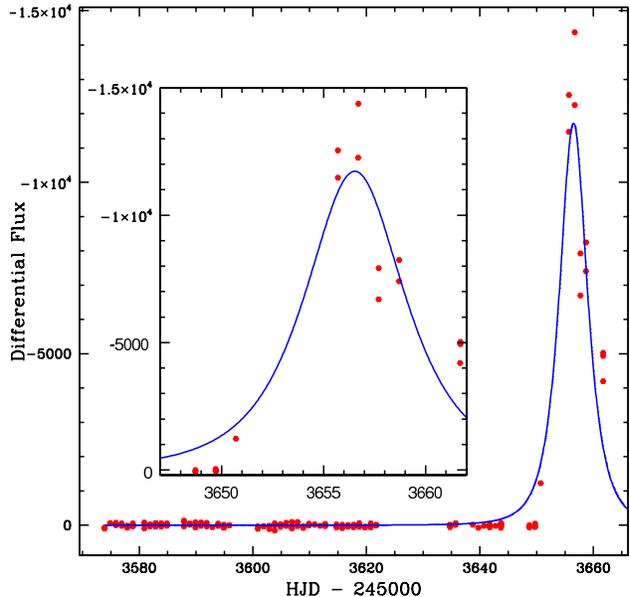}
\end{center}
\caption{An interesting light curve of unknown nature identified through a microlensing search. The counts are negative as an artifact of the subtraction process, the event was in fact a brightening. We estimate a peak magnitude of I$\sim$10.}
\label{BCMLE}
\end{figure}

We did not find any high-confidence, previously undetected
microlensing candidates. We did however indirectly identify a similar
light curve that is interesting in its own right. It was approximately symmetric,
it's time of maximum amplification was several days removed from the
edges of our observation window, and it's timescale was sufficiently
short that a flat baseline is discernible; it is presented with its
model fit in Figure \ref{BCMLE}. The angular position of the event is
$(\alpha, \delta)=$ (17:58:31, -26:54:43).  The closest object found
in a Simbad search was IRAS-17554-2654 at a distance of $45\arcsec$
($\sim$3 pixels), and it is classified as an ``Infra-Red source'',
which is a magnitude 15 object at 12 $\mu$m. It is unfortunately not within sky regions covered by any OGLE field online. We also could not find a match in the ASAS catalog \citep{1997Pojmanski}. The reduced $\chi^2$ for
the fit is very high, 111.5. The best fit parameters were $(\Delta
F_{max}, u_{min}, t_o, t_e)$ = (11720, 1.0, 3656.5, 3.14).

\subsection{Discussion} 

One might ask what the purpose was of using two different methods, if
the second method requires removing sources identified in the first
method? We considered it valuable to compare different approaches. The
first method, where its sources came from a DAOphot list obtained from
the reference image, had the advantage that its relative
fluxes on the difference images could be easily converted to
magnitudes, as well as there being a large volume of proven software
already designed for source identification. The second method has
fewer biases preventing completeness, as variability does not need to
occur within the frames used to construct the reference image. More
empirical study is required to determine the optimal method studying
microlensing curves that are entirely within the high-brightness domain.

The photometric capacity to refine parameters of known light curves is
demonstrated by Figures \ref{KME1} \& \ref{KME2}. The method we
employed is not only effective at performing its task, but relatively
easy to implement. Following proper reductions of the images, one need
only feed ISIS a source list whose coordinates are obtained from the
general surveys. The light curves can then be compared.

\section{Other Variable Stars}
\label{section:OVS}

\begin{figure}[ht]
%\plotone{JstetVsMagFixed100}
\includegraphics[totalheight=0.30\textheight]{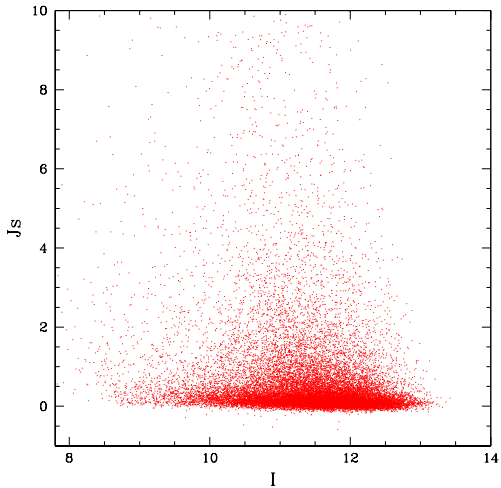}
\caption{Plot of Stetson's ``J'' index ($J_{s}$) of correlated variability vs apparent
magnitude in the I-band for representative sample of points.}
\label{Jstetson}
\end{figure}

Studies of variable stars towards the Galactic bulge have been a
natural and important byproduct of microlensing surveys in that
direction.  For example, OGLE-II published a catalog of $\sim$ 200,000
variable stars discovered in the bulge during its first three years
\citep{2002Wozniak}. Similarly, some 50,000 variable stars
have been detected in that direction by the MACHO project
\citep{1995Cook}. At brighter magnitudes, a survey with similar
equipment to what we have used here - albeit also using a longer
baseline and much higher cadence - would be very effective in
identifying and classifying previously undiscovered variable stars.

With current data, we have light curves for $\sim$ $10^5$ sources, so
we cannot match the sheer number of variable stars discovered by
deeper microlensing surveys. We do however cover a different region of
the parameter space: the brightest stars. We have nearly 800 sources
brighter than magnitude 9, $\sim$ 5,900 sources brighter than
magnitude 10, and $\sim$ 32,000 sources brighter than magnitude 11.

\begin{figure*}[ht]
%\plotone{EclipsingMosaic}
\includegraphics[totalheight=0.82\textheight]{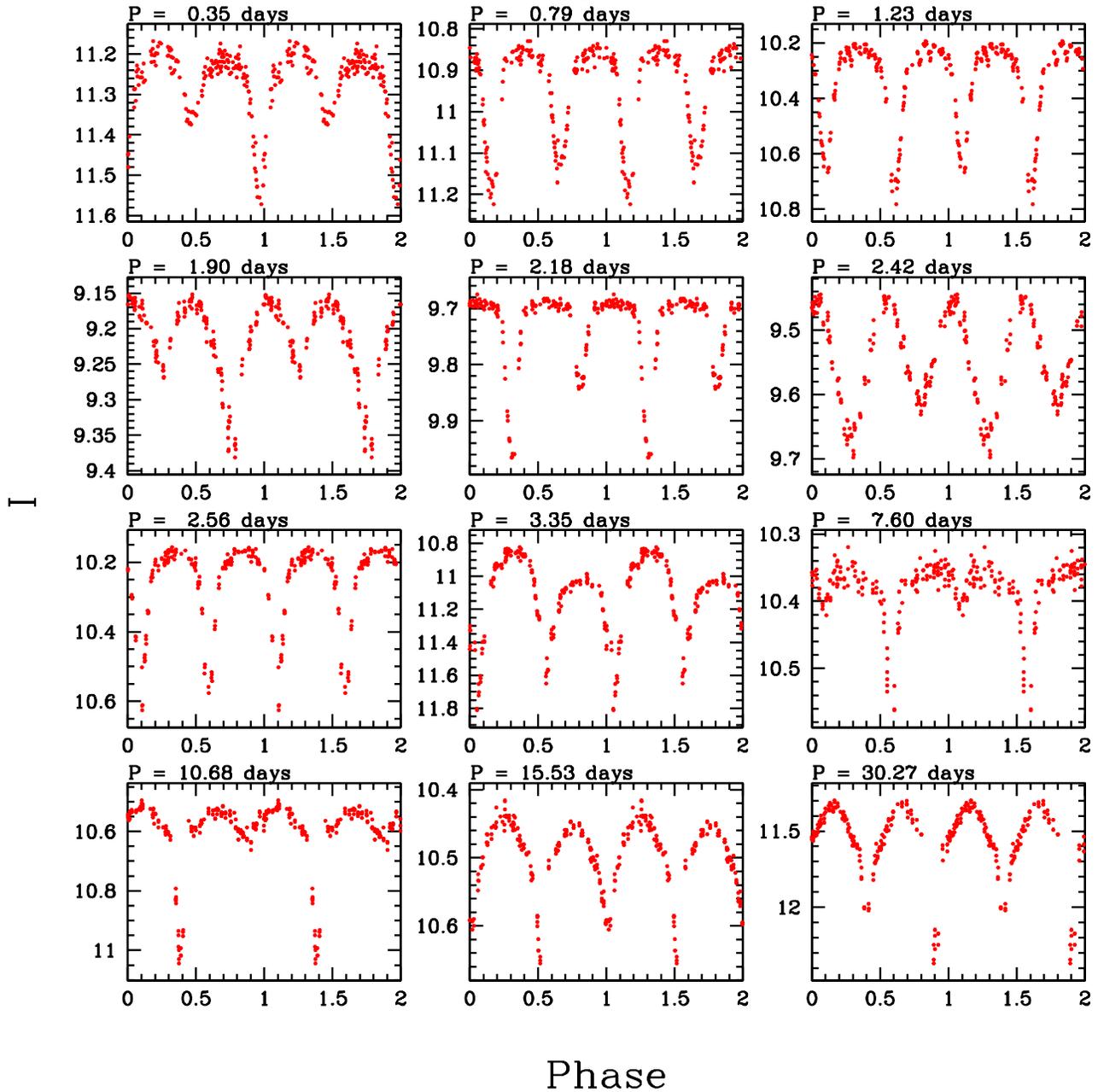}
\caption{Sample of 12 eclipsing variables.}
\label{Eclipsingmosaic}
\end{figure*}

We classify our variables as either periodic or
miscellaneous. Periodic variables were identified using an analysis of
variance search developed by \citet{1989Schwarzenberg} and
\citet{2005Devor}, as implemented by the Vartools package for light
curves \citep{2008Hartman}. On the 66 nights we have observations, the
epochs came within $\sim$30 minutes of each other, so we
effectively have 66 distinct points scattered erratically over the 88
night span of our observations. Aliasing was a significant issue in
this domain, with most stars best-fit periods clustering in a small
subset of frequency space, particularly the harmonics of 1 day. Figure
\ref{Eclipsingmosaic} is a sampling of some of the cleaner eclipsing
variables we found in our data. Figure 9 is in turn a sample of other
periodic variables.

Within this sample alone and using a $30\arcsec$ matching radius, we
find 19 of the 36 variable sources shown here are without \textit{any}
match in Simbad. Expanding the matching radius to $45\arcsec$ yields 4
additional matches. It is reasonable to estimate that around half or
perhaps more of the variables which would show up in a more complete
survey would constitute new discoveries. Further, many of the matches
one would find are with the 2MASS catalog \citep{2003Cutri} and the
IRAS catalog \citep{1986Kleinmann} and thus may contain incomplete
variability information. The 7 matches we found in this manner with information on the period are from the General Catalog of Variable Stars, \citep{1971GCVS}. We summarize those in Table
2. The ASAS catalog \citep{1997Pojmanski} had matches to 6 of our 12 eclipsing variables and 7 of the 12 other periodic variables. As the ASAS and GCVS matches largely overlapped, this left 11 of the 24 sources selected without any match.

\begin{table}[t]
\caption{Periodic variables found to have matches in SIMBAD which include periodicity information. Periods are given in days.}
\scalebox{0.91}
{\begin{tabular}{ l c l l l l l l l}
\hline\hline
Name & Archival & Measured & Class   \\
 & Period & Period & \\
\hline
V* BS Sco &  7.622 & 7.60 & EB (Algol) \\
V* V712 Sco & 30.305 & 30.27 & EB (Algol)\\
V* V3254 Sgr & -- & 3.35 & EB (O'Connell)\\
V* V1188 Sgr & 0.581 & 0.58 & RR Lyr \\
V* V773 Sgr & 5.748 & 5.75 & $\delta$ Cep \\
V* V1828 Sgr & 12.972 & 12.85 & $\delta$ Cep \\
V* V1290 Sgr & 27.9516 & 28.14 & Cepheid \\
\hline
\end{tabular}
\label{table:PeriodicMatches}}
\end{table}

One variable source standing out is the 3.35 day period eclipsing binary as it is demonstrating the O'Connell effect, a light curve asymmetry between outside eclipse maxima, see \citet{1951Oconell} and \citet{1984Davidge}. We estimate the brightness shift between
maxima as $\Delta$I $\approx +0.2$. Located at $(\alpha, \delta)$ =
(18:24:29, -32:25:57), it has a $30\arcsec$ match with V*V3254
Sgr, listed as an eclipsing binary without a specified period,
without the O'Connell effect mentioned and a magnitude offset of
$\sim$ 2.5. 

\clearpage
\clearpage

\begin{figure}[p]
\begin{center}
%\plotone{PeriodicMosaic}
\includegraphics[totalheight=0.82\textheight]{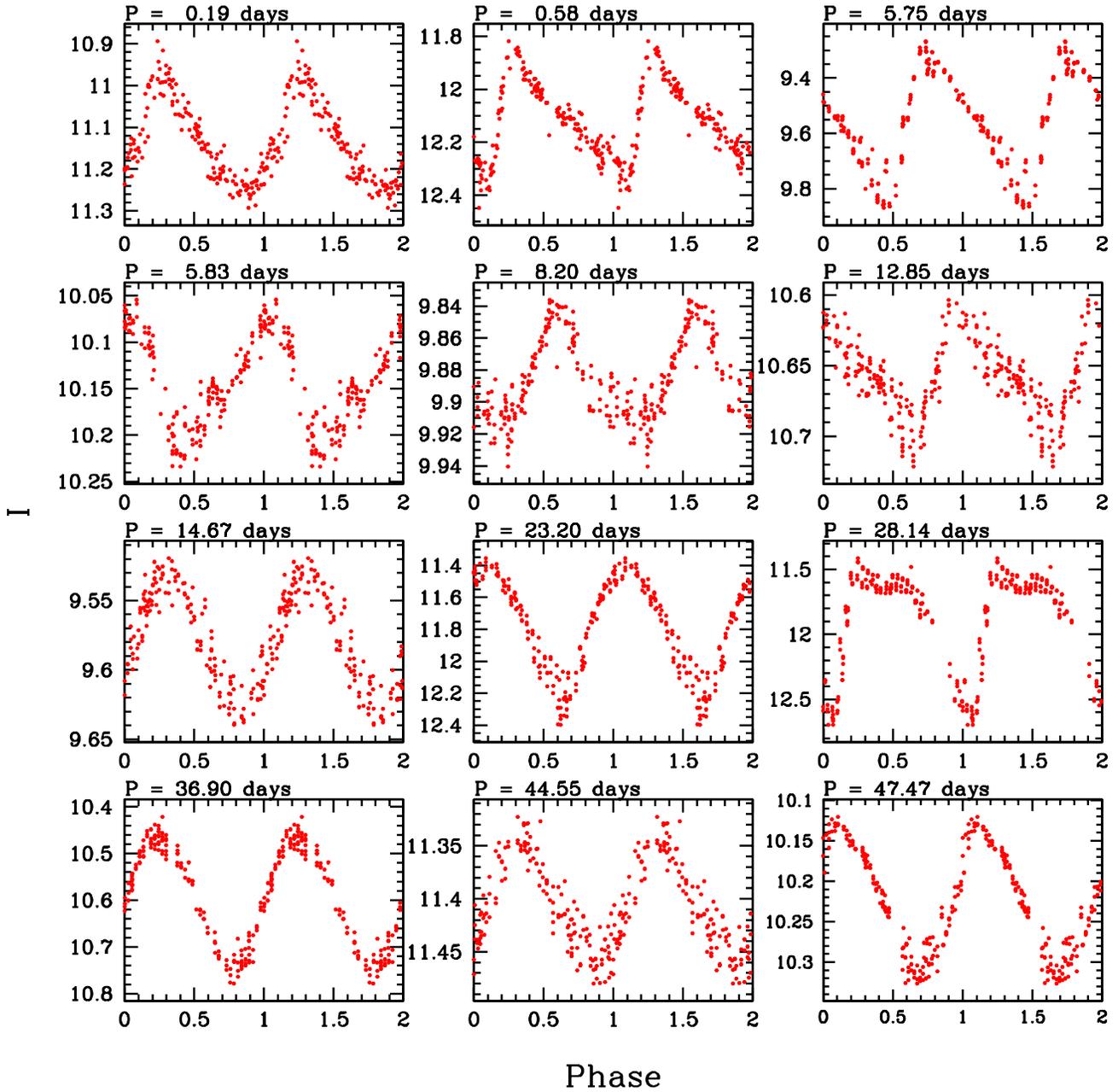}
\caption{Sample of 12 periodic variables.}
\end{center}
\label{Periodicmosaic3}
\end{figure}
\clearpage

\clearpage
\begin{figure*}[h*]
\includegraphics[totalheight=0.82\textheight]{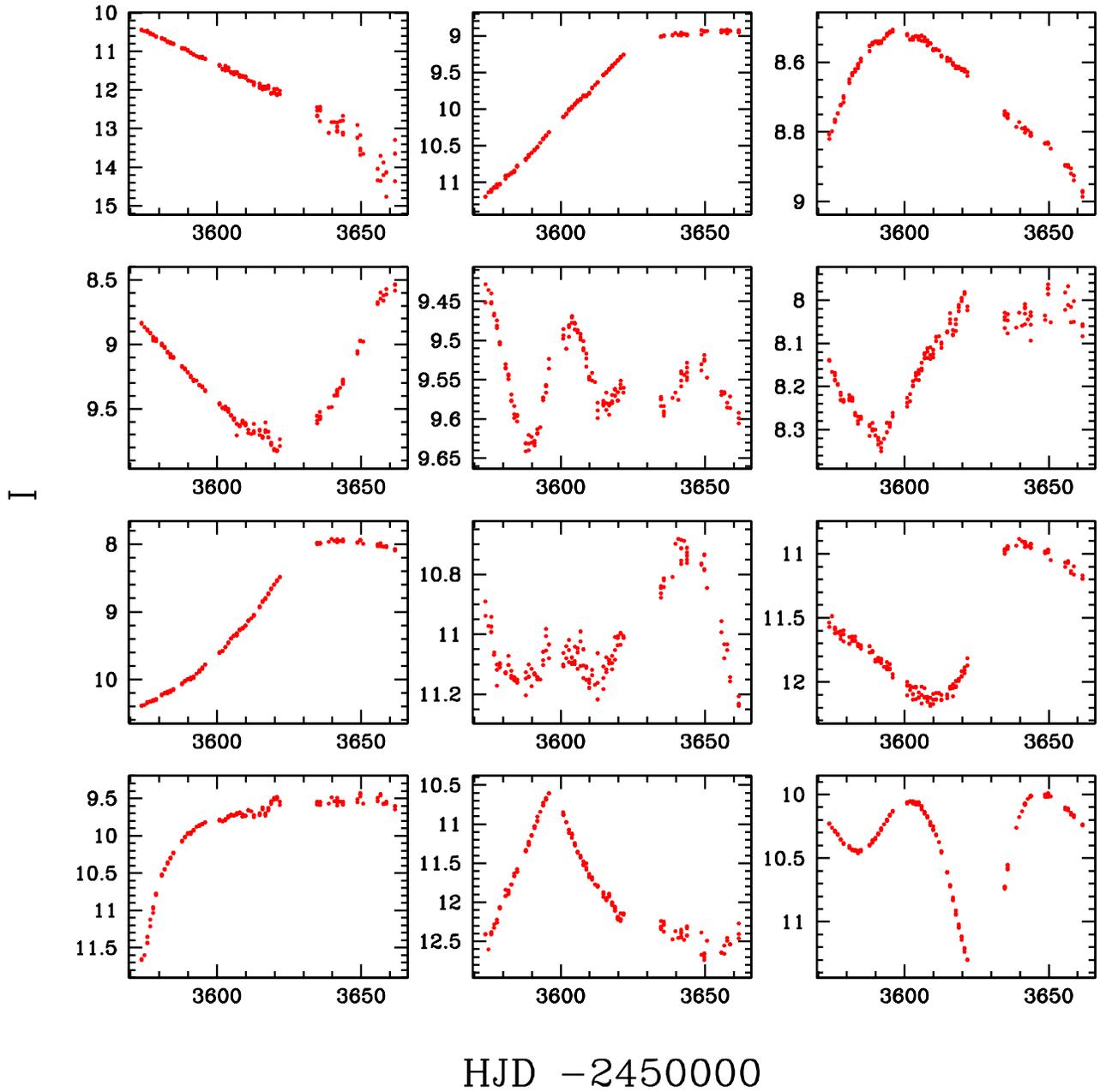}
\caption{Sample of 12 miscellaneous variables.}
\label{LPVmosaic}
\end{figure*}
\clearpage

To search for miscellaneous variable stars we apply a cut of $J_{s}>0.85$,
see Figure \ref{Jstetson}. This selects 23,072 light curves as being
candidate variables. There is nothing intrinsic about the value 0.85, it
was chosen empirically to be where light curves began to appear
unequivocally variable when the time series were viewed by
eye. There likely were true variables below the cutoff, but we
wanted to minimize the risk of false positives, and as such,
sacrificed completeness. A plethora of variables was available at all
signal levels, with 9,775 sources having $J_{s}>2.0$ and 1,264 having
$J_{s}>10.00$. Sources with a high $J_{s}$ value but without a clear
period numbered in the thousands. These were the variable sources we
classified as miscellaneous. A selection is shown in Figure
\ref{LPVmosaic}. We had  $30\arcsec$ astrometric matches in SIMBAD for 9 of these but there was insufficient information to determine if they
are all a correct match.

We choose not to use a larger matching radius for these sources as
without a smoking gun (such as period) it is difficult to determine
the validity of the match. For the 12th source, on the bottom right
corner of Figure \ref{LPVmosaic}, we find the classification scheme is
in fact incorrect. Our match, V* V738 Sgr, is currently classified as a variable star of W Vir type with a period of 43.39 days \citep{1971GCVS}. We find that a 43.39 day period does not fit the light curve. We know that it is the same star as it is well fit by a period of 43.30$\times$2 = 86.78 days. Further data would be required to confirm if this is indeed the correct period.

\section{Conclusion}
\label{section:Conclusion}

Using data from an imaging system with a wide FOV, fast focal ratio,
and large pixel scale, combined with image subtraction photometry, we
have found matches to 7 OGLE microlensing events from the 2005 bulge
season. We searched for previously undetected microlensing events
using two distinct methods, fitting parameters to our DAOphot source
list and fitting parameters to a new source list directly constructed
from the difference images. We found a few events closely resembling
microlensing events but none with a satisfactory fit. In a search for
periodic variables we found that there are a great many
high-brightness eclipsing and periodic variables still to be found in
the bulge. A catalog of periodic variables based on the data discussed
in this paper is in preparation (Nataf et al. 2009).

Buoyed by the proof-of-concept performance achieved here we encourage
the development of Small Aperture Microlensing Survey (SAMS) (see
Table~1). Such a suite of three small aperture telescopes equipped
with high QE, large format CCD cameras would provide high sampling of
the brightest bulge microlensing events along with a more complete
catalog of bright bulge variables. This instrument would naturally
be located in the Southern Hemisphere, and from the logistics point of
view would probably be best located next to existing and planned larger
aperture microlensing telescopes.

\acknowledgements{HATNet operations have been funded by NASA grants
NEG04GN74G, NNX08AF23G and SAO IR\&D grants. We thank J.D. Hartman for
assistance with use of the Vartools package, G. Pojmanski for his
``lc'' program, W. Pych for his ``fwhm'' program, R.J. Siverd for
explanations and help with Perl programming, S. Kozlowski for
suggestions on microlensing fitting, L. Wyrzykowski for his
microlensing curve fitting program, and A. Gould for insights into the
observational process. We made use of the Simbad database, operated in
Strasbourg, France by the \textit{Centre de Données astronomiques de
Strasbourg}.}

%abbrvnat
%unsrtnat
%apalike
%amsplain
\bibliographystyle{abbrvnat}
\bibliography{HatNatafPaper1}

\end{document}